\documentclass[final,5p,times,twocolumn,authoryear]{elsarticle}
\usepackage{soul}
\usepackage{amssymb}
\usepackage{lipsum}
\usepackage{xcolor}

\usepackage{hyperref}
\hypersetup{colorlinks=true, linkcolor=blue, citecolor=blue, urlcolor=blue}
\usepackage{comment}

\defcitealias{2023MNRAS.525..898L}{L23}
\newcommand{\Ly}{\citetalias{2023MNRAS.525..898L}}

\newcommand*{\nl}[1]{\textcolor{black}{#1}}

\newcommand{\hlcolor}[2][yellow]{{#2}}

\newcommand{\rev}[1]{\textcolor{black}{#1}}
\newcommand{\revv}[1]{\textcolor{black}{#1}}

\journal{High Energy Astrophysics}

\begin{document}

\begin{frontmatter}

\title{Average soft X-ray surface brightness profile \\ of massive galaxy clusters in \texttt{Magneticum} simulations}

\author[a]{Aleksei Kruglov}
\author[a]{Natalya Lyskova}
\author[c,d,e,a]{Ildar Khabibullin}
\author[f,g]{Veronica Biffi}
\author[d,e]{Klaus Dolag}

\affiliation[a]{organization={Space Research Institute (IKI)},
                addressline={Profsoyuznaya 84/32}, 
                city={Moscow},
                postcode={117997},
                country={Russia}}


\affiliation[c]{organization={Rudolf Peierls Centre for Theoretical Physics, Department of Physics, University of Oxford, Clarendon Laboratory},
                addressline={Parks Rd},
                city={Oxford}, 
                postcode={OX1 3PU},
                country={United Kingdom}}

\affiliation[d]{organization={Universit\"ats-Sternwarte, Fakult\"at f\"ur Physik, Ludwig-Maximilians-Universit\"at M\"unchen},
                addressline={Scheinerstr.~1}, 
                city={M\"unchen},
                postcode={D-81679}, 
                country={Germany}}  

\affiliation[e]{organization={Max Planck Institute for Astrophysics},
                addressline={Karl-Schwarzschild-Str. 1}, 
                city={Garching},
                postcode={D-85741},
                country={Germany}}    
                
\affiliation[f]{organization={INAF, Osservatorio Astronomico di Trieste},
                addressline={via Tiepolo 11}, 
                city={Trieste},
                postcode={I-34131}, 
                country={Italy}}     

\affiliation[g]{organization={IFPU – Institute for Fundamental Physics of the Universe},
                addressline={via Beirut 2}, 
                city={Trieste},
                postcode={I-34014}, 
                country={Italy}}     

\begin{abstract}
The self-similar growth of massive galaxy clusters suggests that radial profiles of their key thermodynamic properties should have identical shapes after proper mass- and redshift-dependent re-scaling. This property, tested within the virial radius on samples of well-studied individual objects, together with clear and robust observational characteristics such as sensitivity and background accounting, enables the possibility of stacking observations that can be confronted with identically-derived population-averaged predictions from theory or numerical simulations at large radii. Such a comparison not only eliminates effects of inevitable stochasticity in properties of individual objects, but also allows one to reach higher sensitivity for the faintest regions in the outskirts of the clusters. In this study, we conduct a one-to-one comparison of the observed and simulated average soft X-ray surface brightness profiles of several dozen massive galaxy clusters at low redshift. We find a very good out-of-the-box agreement between the $0.3 - 2.3$ keV surface brightness profile of stacked galaxy clusters recently measured by \textit{SRG}/eROSITA and the corresponding predictions from the \texttt{Magneticum} cosmological hydrodynamical simulations, which are known to reproduce other scaling relations observed for massive galaxy clusters. A significant difference between the observed and simulated profiles is present in the very central region, where effective implementation of the AGN feedback likely results in excessive gas redistribution within the core. The simulations predict a very noisy surface brightness profile beyond several times the virial radius of the cluster, with the mean signal being orders of magnitude lower than the local radially-flat but strongly fluctuating emission background, meaning that a proper detection of this component would be very challenging even with larger samples in the future.
\end{abstract}


\end{frontmatter}

\section{Introduction}
\label{intro}

Galaxy clusters are the largest gravitationally bound structures in the observable Universe, with over 99 percent of their mass composed of dark matter and hot X-ray–emitting gas \citep[for a review]{2012ARA&A..50..353K}. They form as a result of the continuous mergers, and these processes are most noticeable in the peripheral regions, where its current growth occurs through the inflow of matter from the surrounding environment \citep{2007PhR...443....1M}. However, the density and temperature of hot gas and, as a result, the intensity of radiation are much lower at the periphery of the cluster than at its center, and, moreover, gas is far from hydrostatic equilibrium in the outskirts \citep{1996ApJ...469..494E, 2013SSRv..177..195R}; nevertheless, there are detailed studies of the outskirts in particular cases \cite[e.g.][]{2024A&A...681A.108V, 2023A&A...670A.156C, 2026A&A...707A.381C}. In addition, the individual features of clusters make it difficult to generalize the characteristics of a single object to the entire population. One way to solve these problems is to combine and average data from many clusters, which allows to expand the radial range of X-ray detection by reducing statistical errors \citep{2012A&A...541A..57E, 2021MNRAS.504.4649O, 2022MNRAS.514.1645A}. This is especially important considering that in a self-similar framework \citep{1986MNRAS.222..323K}, the shapes of the galaxy clusters thermodynamical radial profiles, for example, those of gas density and temperature, demonstrate approximately universal behaviour after appropriate scaling, i.e., by radius or mass \citep{2002A&A...389....1A, 2007ApJ...668....1N,2014ApJ...789....1D, 2015ApJ...806...68L, 2019A&A...621A..41G, 2022A&A...665A..24P, 2024MNRAS.533.2656B}. The averaged cluster is expected to exhibit spherical symmetry, as stacked and averaged radial profile smooth out features caused by anisotropic growth of individual clusters.

Recently, with the launch of the \textit{Spectrum-RG} observatory \citep{2021A&A...656A.132S} and completion of several all-sky scans with \textit{SRG}/eROSITA telescope \citep{2021A&A...647A...1P}, it has become possible to observe individual clusters with essentially ``unlimited'' field of view, allowing for proper stacking analysis of the X-ray data beyond radius $R_{200c}$\footnote{Throughout this paper we will use following notation: $R_{200c} \ (R_{500c}$) and $R_{200m}$ correspond to the radii within which the mean density is equal to 200 (500) times the critical density and 200 times the mean density of the Universe at the redshift of the cluster, respectively. Hereafter, $R_{500c}$ is denoted as $R_{500}$.}, to which most cluster outskirts studies are limited. \cite{2023MNRAS.525..898L} used eRASS (eROSITA All-Sky Survey) data to derive the stacked X-ray surface brightness and gas density profiles for a sample of galaxy clusters selected from the CHEX-MATE catalogue \citep{2021A&A...650A.104C}: the cluster emission was detected up to $\sim 3 \times R_{500c}$, where the surface brightness of the stacked image drops below $1\%$ of the background. Moreover, \cite{2026A&A...709A..72Z} stacked 680 low-redshift clusters selected from eRASS survey \citep{2024A&A...685A.106B} and detected a statistically significant excess X-ray signal above the background at the $(1-2)\times R_{200m}$.

To place the results of \cite{2023MNRAS.525..898L} (\Ly~hereinafter) in a theoretical context, we reproduce the same pipeline for a comparable sample of simulated galaxy clusters. In this work, we study the hot gas distribution in massive (in particular, $>10^{14} M_{\odot}/h$) clusters from the \texttt{Magneticum}
suite of state-of-the-art cosmological hydrodynamical simulations \citep{2013MNRAS.428.1395B, 2014MNRAS.442.2304H,2016MNRAS.463.1797D, 2025arXiv250401061D}. We perform mock observations of galaxy clusters modelled for the parameters of the all-sky survey by \textit{SRG}/eROSITA, stack and average their X-ray images, derive mean surface brightness profile and compare it with the data from real observations.

The paper is structured as follows: we describe the generation and filtering of cluster images in Section~\ref{sec:methods} and present the surface brightness profile of the stacked image in Section~\ref{sec:profiles}. In Section~\ref{sec:discussion}, we discuss the results and  summarize our findings in Section~\ref{sec:conclusions}. \hlcolor{Throughout the paper, we assume a $\Lambda$CDM cosmology with $\Omega_{\Lambda} = 0.728, \ \Omega_{\rm M} = 0.272 \ (16.8\%$ baryons) and $H_0 = 70.4$ km/s/Mpc\rev{, i.e. $h=H_0/(100$~km/s/Mpc$)=0.704$}.}

\section{Methods}\label{sec:methods}

\subsection{Cluster sample}\label{sec:sample}

\begin{figure}
  \centering
  \includegraphics[width=0.95\linewidth,trim=0cm 0.5cm 0cm 0cm]{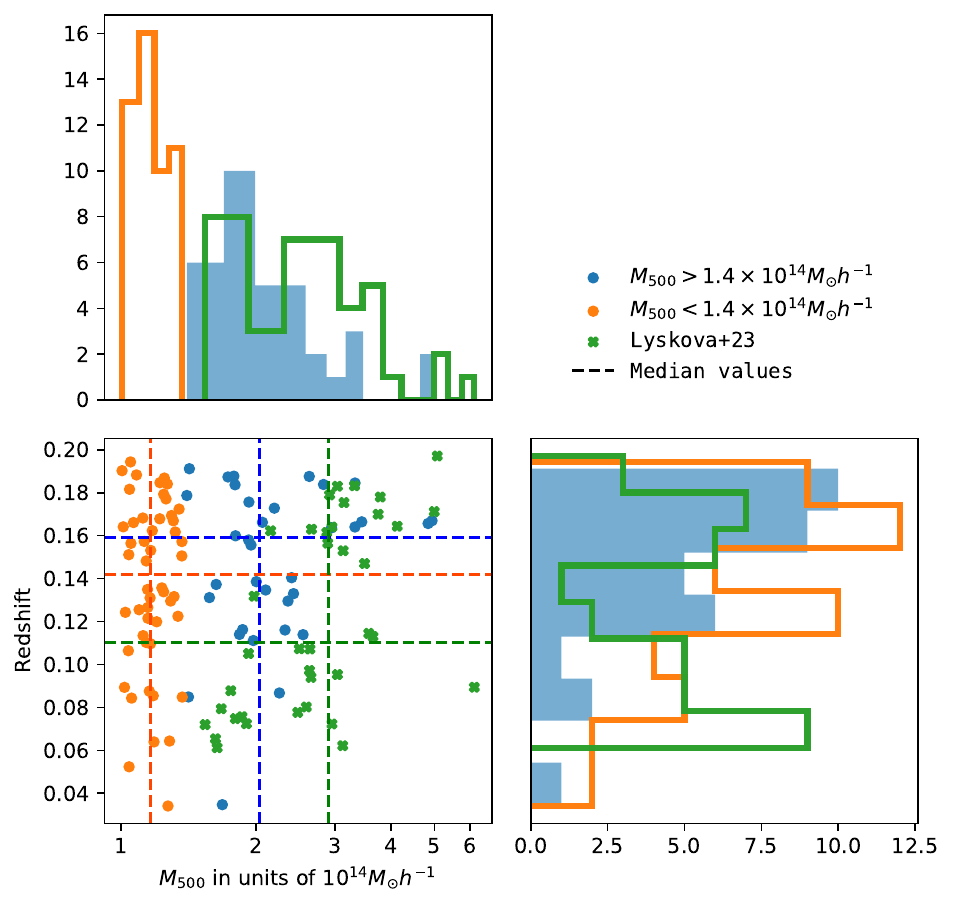}
  \caption{Histograms of the masses $M_{500}$ and redshifts $z_{\rm true}$ for two subsets of clusters in our sample (50 clusters with $M_{500}$ below $1.4 \times 10^{14} M_{\odot}/h$ in orange, and 34 clusters with $M_{500}$ above this value in blue) and 38 clusters used for SRG/eROSITA stacking in \Ly~ in green. Dashed lines show median values for each subset.}
  \label{fig:histograms}
\end{figure}

The main dataset used for this study is the same as in \cite{2025JCAP...05..007K}. Briefly, it is based on a $30 \times 30$ degrees lightcone produced by means of the \texttt{PHOX} software package \citep{2012MNRAS.420.3545B,2013MNRAS.428.1395B}. This particular lightcone was drawn  from the \texttt{Magneticum} \textit{Box2/hr} simulation and includes 84 massive clusters with masses $M_{500}$ above $10^{14} M_{\odot}/h$. \rev{The most relevant resolution parameters of the \textit{Box2/hr} simulation for this study are the gas particle mass $m_{\rm gas} = 1.4\cdot10^8 M_{\odot}/h$ and the gravitational smoothing length $\epsilon_{gas} = 3.75\ {\rm kpc/h}$, implying that the gravitational resolution limit is $\sim15$ kpc/h} \citep{2025A&A...694A.232G}\rev{; this ensures that resolution effects do not affect any drawn predictions beyond the most central part of the clusters, i.e. a few percent of the $R_{500}$.} More details about \textit{Box2/hr} setup parameters and lightcone geometry, as well as description of synthetic observations of galaxy clusters from this particular lightcone in the context of \textit{SRG}/eROSITA survey are available in \cite{2023A&A...670A..33S, 2023A&A...675A.150Z, 2024A&A...689A...7M, 2025arXiv250401061D}. The properties of galaxy clusters and groups from the lightcone are summarized in a catalogue that is made publicly available (see the data section of the \texttt{Magneticum} project webpage) and to which we will refer throughout this paper. We divide the sample of 84 simulated clusters into two subsets \hlcolor{using a mass threshold of} $M_{500} = 1.4 \times 10^{14} M_{\odot}/h$ (which is close to the lower mass cut of the sample from \Ly)\hlcolor{ and show the distributions of masses $M_{500}$ and redshifts for each subset in} Fig.~\ref{fig:histograms}. \hlcolor{In the following, we derive our results using the sample of 34 clusters with $M_{500} > 1.4 \times 10^{14} M_{\odot}/h$.}

\rev{The subsample of CHEX-MATE used in \Ly~is considered representative of the general cluster population, as is the \texttt{Magneticum} sample, which contains all clusters above the corresponding mass threshold in the 352 Mpc/h box. Hence, we do not expect any significant systematic differences in their properties that would introduce substantial biases in our comparison. In particular, the cool-core fraction reported for CHEX-MATE is approximately 40--50\%, with the precise value depending on the adopted definition \citep{2022A&A...665A.117C,2026ApJ..1000..252W}. This is consistent with the cool-core fraction found for the corresponding \texttt{Magneticum} sample \citep{Hinz2018,2025A&A...694A.232G}.}

\subsection{X-ray images and filtering}\label{sec:images}

For each cluster, we collect all photons emitted by the gas elements (assuming optically thin thermal emission) in the $0.3-2.3$ keV range within a cylinder with a radius equal to $10 \times R_{500}$. \rev{The predicted X-ray emission from AGN in individual galaxies is also available \cite[]{2018MNRAS.481.2213B}, but not included in our analysis to ensure consistency with \Ly.} \rev{The contribution from the unresolved AGN emission to the cluster surface brightness profile is expected to be small \cite[e.g.][]{2004A&A...419...47F, 2009ApJ...692.1033V}}.
Only photons from the lightcone slice corresponding to the cluster redshift are included, excluding contributions from other slices. Nevertheless, the line-of-sight thickness of each slice \hlcolor{($\delta z\approx0.033$, corresponding to $\gtrsim100$ Mpc)} ensures that correlated large-scale structures surrounding these massive clusters are incorporated into the extracted photon lists due to projection effects.

For each simulated photon, we calculate the corresponding effective area using the FoV-averaged ARF \hlcolor{(\textit{Auxiliary Response File})} appropriate for the all-sky survey observations of \textit{SRG}/eROSITA (e.g. \Ly), and use this value as a weight when generating the image. We set the center of each cluster at the brightest pixel of the Gaussian-smoothed ($\sigma=40''$) image. The side of an image is equal to $20 \times R_{500}$. For each cluster, its image has a resolution of several arcseconds per pixel (ranging from $1.4''$ to $6.9''$). The values in each bin are rescaled to represent the surface brightness: they are divided by $1000 \ \rm{cm}^2$ (flat effective area for photons), $10^4$ s (nominal exposure time for photons) and the area of the bin in arcmin$^2$. For subsequent stacking, the images are re-mapped as if, for each cluster, $R_{500}$ were equal to 1 Mpc (or 10 arcmin) (i.e. the surface brightness is also corrected according to  Eq. B4 in \Ly). Finally, the images are smoothed to match the \textit{SRG}/eROSITA FoV-averaged PSF half-energy width \hlcolor{(i.e. ${\rm FWHM} = 26''$ and $\sigma={\rm FWHM}/{2.355} = 11''$)}. We  demonstrate the effect of different PSFs on the surface brightness profiles of stacked images in \ref{app:psf}.

Since almost every simulated individual image includes significant number of bright substructures, we perform a masking and filtering procedure as follows. For each cluster's image we \hlcolor{identify} all less-massive haloes from catalogue that belong to the same $20 \times 20 \ R_{500}$ field (i.e. not only from the same slice, but from all the slices). To remove the contribution of these haloes, we produce a mask (exposure map) from which the circles with radii equal to $R_{200c}$ for each halo are excluded, but leaving the area within $R_{500}$ of considered cluster unfiltered. However, in several cases this filtering does not properly remove all sources in the vicinity. Either for bright substructures \hlcolor{(i.e. "pockets" of dense and relatively cold gas)} there is no corresponding halo in the catalogue or the halo is too extended to be properly removed by the method described above. Moreover, it is possible that considered cluster demonstrates (kind of) dynamical behaviour and has very close ($\gtrsim R_{200c}$) but distinct satellites. Thus, for \hlcolor{some} clusters the manual filtering have been applied during which extra-bright regions were removed (i.e., exposure time for these areas was set to zero). \hlcolor{Catalogue indices of specific manually filtered clusters} are provided in \ref{app:manual} with example images. To avoid overfiltering the images, we apply the following criterion: substructures are removed only until the average profile no longer shows any change. 

\begin{figure}
  \centering
  \includegraphics[width=\linewidth,trim=0cm 0.5cm 0cm 0cm]{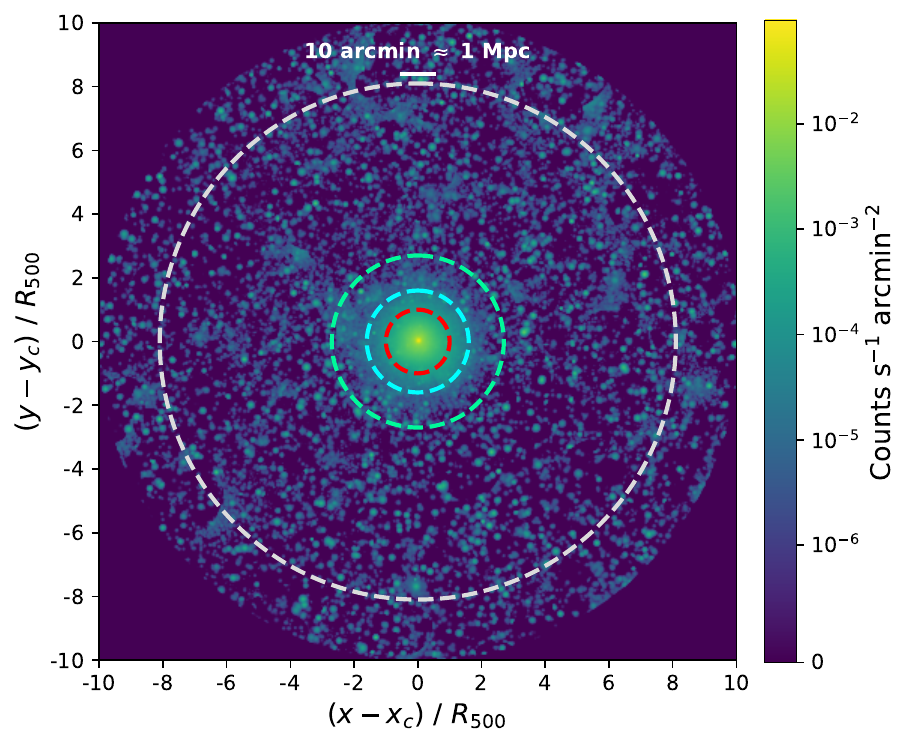}
  \caption{Stacked exposure-corrected 0.3-2.3 keV image of 34 clusters with filtering of extended sources. The image is $20 \times R_{500}$ on a side. The dashed circles show positions of $R_{500c}$, $R_{200c}$, $R_{200m}$, and the turn-around radius $R_{ta}$. We adopt the following relation between these radii $R_{ta}:R_{200m}:R_{200c}:R_{500c} = 8.1:2.7:1.6:1$ \citep{2014ApJ...789....1D,2014ApJ...792...25N}.}
  \label{fig:stacked_image_34}
\end{figure}

To obtain the stacked image we sum the surface brightness values in each pixel over all images, sum the exposure maps and divide the former by the latter. The stacked X-ray image is shown in Fig.~\ref{fig:stacked_image_34}.

\section{Results}\label{sec:profiles}

\begin{figure}[!ht]
  \centering
  \includegraphics[width=0.99\linewidth,trim=0cm 0.88cm 0cm 0cm]{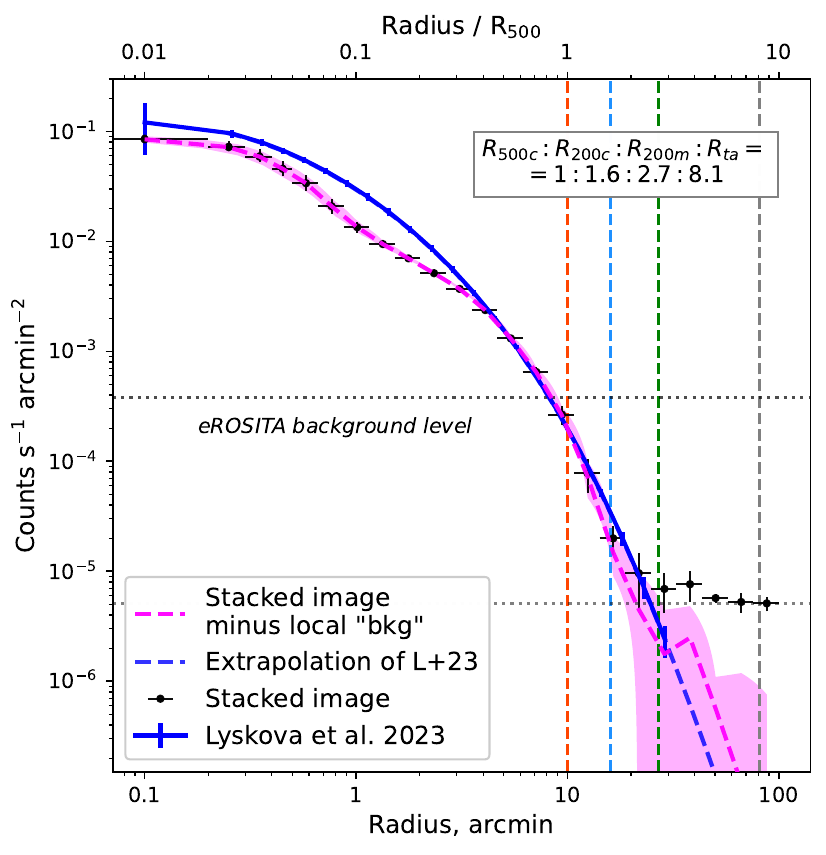}
  \hspace*{0.089cm}
  \includegraphics[width=0.968\linewidth,trim=0cm 0.3cm 0cm 0.15cm, clip]{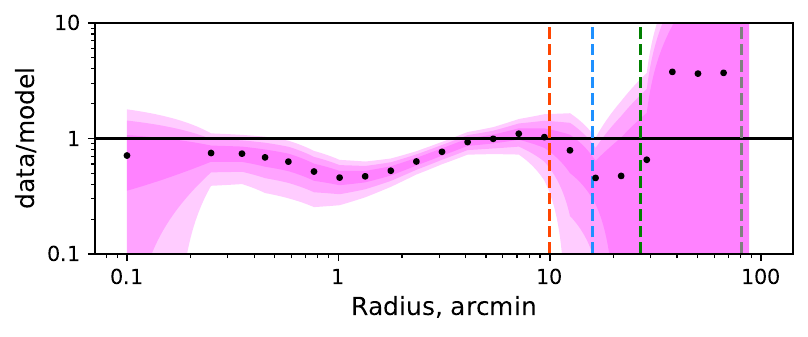}
  \caption{
  \rev{\rev{Comparison of the simulated and the observed radial profiles of X-ray emission.} \textit{Top:}} the radial surface brightness profile of the stacked image in the 0.3-2.3 keV band (black points with errorbars). The \rev{solid blue} line indicates the stacked profile from \Ly~\rev{and the dashed blue line shows the extrapolation of the best-fit model;} the upper dotted line shows the background level measured by \textit{SRG}/eROSITA from that work. \rev{The lower dotted line marks the "local background" level estimated 
  from the values of the surface brightness profile beyond $2 \times R_{200m}$.} 
  The magenta dashed line shows the same profile after subtraction of this constant local background level\rev{, with the $1\sigma$ uncertainty indicated by the magenta-shaded region.} 
   \rev{
   \textit{Bottom:}
  surface brightness profile from the stacked image after subtraction of the local background, divided by values of the extrapolated model from \Ly, with the $1, 2 \ {\rm and} \ 3\sigma$ uncertainties indicated by the magenta-shaded regions.
   }
   }
  \label{fig:profiles_log}
\end{figure}

The surface brightness profile extracted from the stacked X-ray image is shown in \rev{the upper panel of} Fig.~\ref{fig:profiles_log} together with the profile measured by \Ly~\rev{and the extrapolation of the best-fit model from the same work. We calculate the standard deviation $\sigma$ of the surface brightness across four 90-degree wedges to estimate the azimuthal scatter in each radial bin.} Remarkable agreement \rev{within $2\sigma$} is observed across a broad range of radii \rev{($0.4-3 \times R_{500}$, see lower panel of} Fig.~\ref{fig:profiles_log}) and more than two orders in surface brightness, with noticeable exception for the very central region, $R\lesssim 0.4 \times R_{500}$\rev{, where our profile is systematically below the observed profile, with the difference reaching up to a factor of $2$ at $\sim 0.1 \times  R_{500}$}.  At large radii, the profile reaches an approximately flat, background-dominated plateau, which we further define as the ``local background''. \rev{It includes emission from correlated surrounding structures, warm–hot intergalactic medium (WHIM) and residual substructures; in contrast to the actually measured  \textit{SRG}/eROSITA background level, it does not include instrumental/particle background, cosmic X-ray background and Galactic foregrounds like the Local Hot Bubble and Milky Way halo.} We estimate \revv{the level of local background} from the mean surface brightness measured at $R\gtrsim 2 \times R_{200m}$: $(5\pm1) \times 10^{-6}$ counts/s/arcmin$^2$\rev{, which is about 2 orders of magnitude lower than the observed background level of the \textit{SRG}/eROSITA stacked profile ($3.8 \times 10^{-4}$ counts/s/arcmin$^2$).} \rev{This constant level is then subtracted from the stacked profile, allowing} us to trace the \rev{signal} to radii beyond $3\times R_{500}$\rev{, although with very large uncertainties} (see the magenta dashed line in Fig.~\ref{fig:profiles_log}). 

\begin{figure}[ht]
  \centering
  \vspace*{0.31cm}
  \includegraphics[width=\linewidth,trim=0cm 0.9cm 0cm 0cm]{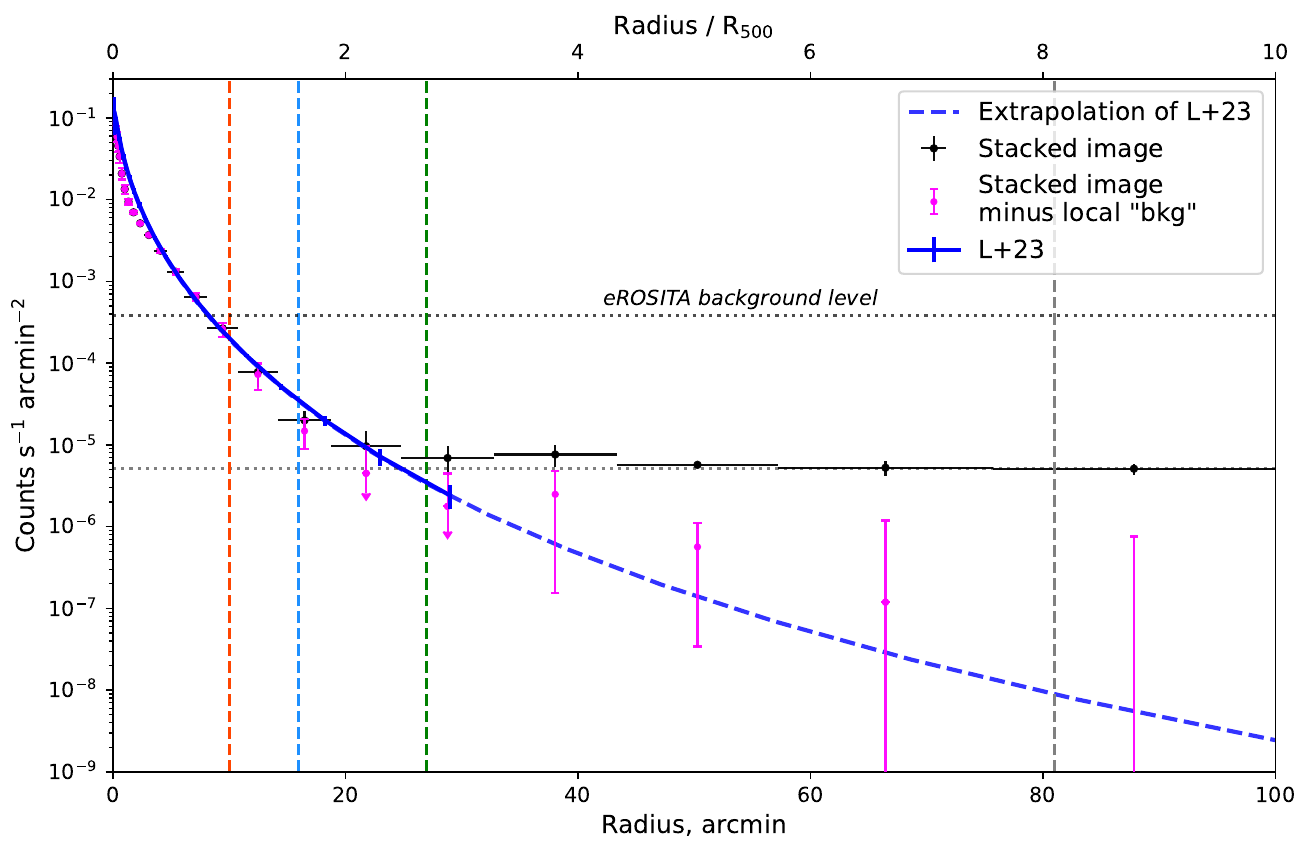}
  \hspace{2cm}
  \includegraphics[width=\linewidth,trim=0cm 0.5cm 0cm 0cm]{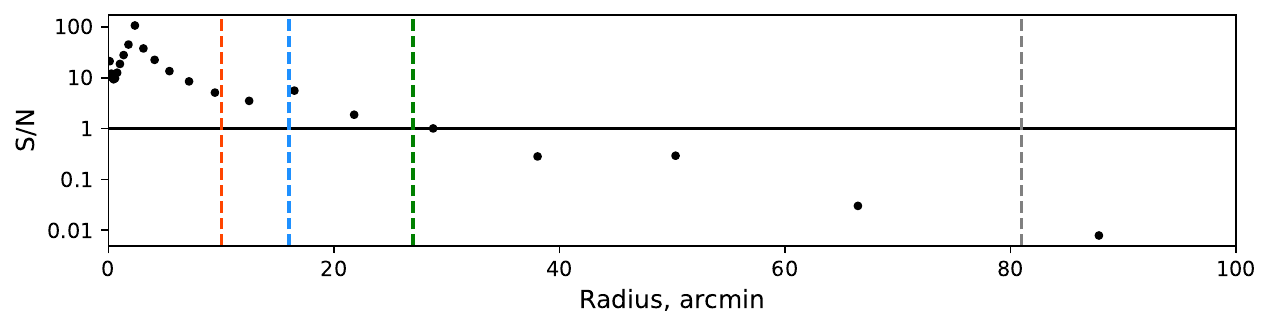}
  \caption{Comparison of the simulated and the observed radial profiles of  X-ray emission.
  \textit{Top:} radial surface brightness profile of the stacked image in the 0.3-2.3 keV band (black points with error bars).
  The solid blue line shows the profile from \Ly~and the dashed blue line shows the extrapolation of the best-fit model from the same work. 
  The upper dotted line indicates the background level measured by \textit{SRG}/eROSITA (from \Ly), while \rev{the lower dotted line marks the "local background" level estimated from the values of the surface brightness profile beyond $2 \times R_{200m}$.}
  The magenta points indicate the surface brightness after subtraction of this constant local background level (plotted with asymmetric error bars).
  The vertical dashed lines mark characteristic radii, including $R_{500c}$, $R_{200c}$, $R_{200m}$, and the turn-around radius $R_{ta}$.
  \textit{Bottom:}
  values of the extrapolated model from \Ly~divided by the surface brightness errors of the stacked image, as a function of radius (signal-to-noise ratio).
  }
  \label{fig:profiles_linear} 
\end{figure}

We also compare the obtained profile with the best-fit model of \Ly~{(see their Appendix C and the corresponding corrigendum)} extrapolated to $10 \times R_{500}$. In order to highlight \hlcolor{the profile behaviour in the outskirts}, we show the same profile on a linear radial scale in Fig.~\ref{fig:profiles_linear} along with the corresponding plot for the signal-to-noise (S/N) ratio. Overall, the stacked emission remains clearly traced up to \hlcolor{$\sim R_{200m} \approx 3\times R_{500}$}, after which the signal-to-noise ratio drops below unity and the profile becomes increasingly background-dominated.

\section{Discussion}\label{sec:discussion}

There is good agreement between the \texttt{Magneticum} stacked surface brightness profile and the \Ly~measurement over $(0.4-3) \times R_{500}$, suggesting that the average radial gas distribution in the \texttt{Magneticum} simulations is broadly consistent with direct X-ray observations. 
\hlcolor{Additionally, we compare our stacked surface brightness profile with that derived directly (i.e. bypassing mock X-ray observations) from the gas density of massive \texttt{Magneticum} clusters presented in} \cite{2022A&A...663L...6A} \hlcolor{and find very good alignment in the local background level. This procedure also allows us to estimate the level of gas clumpiness in the simulated clusters, which we find to be relatively moderate. For details, see}~\ref{app:a22}).

\subsection{Central discrepancy}\label{sec:cd}

\rev{There is a noticeable deviation of the simulated surface brightness profile compared to the observed one in the central part (at $\lesssim 0.4 \times R_{500}$). Several effects could, in principle, contribute to this discrepancy: strong sensitivity of the cluster core to baryonic physics, especially gas cooling and re-heating via AGN feedback; assumed metal enrichment and X-ray emissivity in simulations; differences in the cluster properties in the simulated and observed samples.}

\rev{Among these possibilities, an imperfect treatment of AGN feedback is the most plausible explanation. Recent XRISM measurements of ICM velocity dispersions in cool‑core clusters are systematically lower than those predicted by several state‑of‑the‑art cosmological simulations, indicating that AGN feedback in these models is too strong in the core \citep{2025ApJ...993L..11X, 2026ApJ..1000...75G}. This interpretation is broadly consistent with results from \texttt{Magneticum}: AGN feedback can significantly reshape cool-core structure and suppress the central gas density \citep{2025A&A...694A.232G}. }

\rev{Uncertainties in the assumed metallicity and in the conversion of simulated gas properties into soft-band X-ray emission (i.e. its implementation in the \texttt{PHOX} software package) are unlikely to explain the amplitude of the observed discrepancy.} \rev{In \texttt{Magneticum}, most of the metals currently present in the ICM were produced and dispersed at early epochs ($z>2$) \citep{2023A&A...675A.188A}. Consequently, at the low redshifts considered here, imperfections in modelling of the ICM gas metallicity are expected to be minor, with only a limited effect on the soft-band emissivity. In \texttt{PHOX}, variations in metallicity values can alter the emission measure and, consequently, the surface brightness, by only a few per cent \citep{2012MNRAS.420.3545B}, which is unlikely to account for the observed central discrepancy.} 

\rev{Besides that, the central regions of surface brightness radial profiles may also be affected by the presence or absence of a prominent cool core \citep{2008A&A...483...35S, 2011A&A...526A..79E}; which is less likely to be the reason in our case, given that simulated and observed samples are consistent regarding the fraction of cool-core clusters (see Section~\ref{sec:sample}).} 

\subsection{Towards the outskirts}\label{sec:outskirts}

At large radii ($\gtrsim R_{200m}$), the profile flattens into a background-dominated plateau, which almost certainly \hlcolor{is not associated with the cluster emission} and includes a contribution from correlated surrounding structures (i.e. essentially representing the two-halo term of the correlation function), \hlcolor{warm–hot intergalactic medium (WHIM), diffuse gas, gas clumps, unresolved substructures and far outskirts of substructures and surrounding haloes that were resolved and masked.} \revv{As estimated in Section~\ref{sec:profiles}, this ``local background'' is almost two orders of magnitude fainter than the actual  \textit{SRG}/eROSITA background, making it extremely difficult to detect and robustly subtract in real observations.} \cite{2026A&A...709A..72Z} have shown that proper modelling of the two-halo term and its subtraction significantly reduces residuals in the outskirts, \hlcolor{which, however, is a challenging task due to the extreme faintness and spatial extent of the corresponding signal.} After subtracting the constant ''local background'' level, \hlcolor{the stacked profile remains compatible} with the extrapolation of the \Ly~model up to $3 \times R_{500}$. Beyond this radius, the signal-to-noise ratio drops below unity, and the derived profile in these far outskirts should be interpreted with caution, since it becomes increasingly sensitive to the residual background fluctuations.

\subsection{Increasing of the sample size}\label{sec:largesamples}

The stacked profile for our sample of 34 massive (with $M_{500} > 1.4 \times 10^{14} M_{\odot}/h$) clusters shows very similar behaviour to all 84 clusters (with $M_{500} > 10^{14} M_{\odot}/h$) from the \texttt{Magneticum} lightcone, as well as to the 50 remaining clusters, \hlcolor{implying a self-similar gas distribution over the mass range of the subsamples considered} (see \ref{app:bigsample}). 
\rev{However, increasing the sample from $\sim 30$ to $\sim 80$ clusters does not raise the signal-to-noise ratio above unity at $R \gtrsim R_{200m}$, and detecting cluster emission at these radii remains challenging. As discussed in \Ly, extending the detection radius of cluster emission from $3 \times R_{500}$ to $4 \times R_{500}$ requires increasing the sample size by at least an order of magnitude. Consistently with this estimate, \cite{2026A&A...709A..72Z} used a sample $\sim 20$ times larger and reported a significant\footnote{\rev{Note that \cite{2026A&A...709A..72Z} adopts a different definition of the uncertainty on the stacked surface brightness profile, estimating it from pixel-wise bootstrap resampling, whereas in our analysis the uncertainties are estimated from the scatter among four 90-degree sectors. Adopting different definitions of the uncertainty may lead to different detection significances. In Appendix A of \cite{2026A&A...709A..72Z}, the authors compare the stacked surface brightness profile with profiles measured in four independent 90-degree sectors. The scatter among the sector profiles is substantial, indicating that sector-to-sector variations become important at  $R \gtrsim R_{200m}$, consistent with our results.}} signal from stacked clusters beyond $3 \times R_{500}$.} 

\subsection{\revv{Comparison of stacked and median profiles}}\label{sec:stvsmed}



\nl{Median surface brightness (or other physical quantity) profiles are often used in statistical analyses to characterise the typical properties of a galaxy cluster population. For example, the median X-ray surface brightness profiles of galaxy clusters from several cosmological hydro simulations, for instance \texttt{IllustrisTNG} \citep{2026A&A...707A.122G} and \texttt{CRK-HACC} \citep{2025arXiv251121921F}, have shown a good agreement with observed X-ray profiles obtained through stacking in \Ly. \cite{2023A&A...674A.179B} computed the median profiles for the CHEX-MATE sample and \texttt{THE THREE HUNDRED} simulations and did not reveal a similarly strong systematic offset in the central region ($ \lesssim 0.4 \times R_{500}$). However, median and stacked profiles do not necessarily exhibit the same behaviour, particularly for heterogeneous cluster samples. Here, we investigate possible differences between the median and stacked surface brightness profiles for our sample, and qualitatively examine the relative contributions of azimuthal and cluster-to-cluster variations. }

\nl{To do so, we extract surface brightness profiles in four $90^\circ$ wedges from each individual cluster image (34 clusters in total). This yields 136 profiles, which we treat as individual measurements. We then calculate the median surface brightness profile and the standard deviation (corresponding to the 68\% encompassing quantiles) across all 136 measurements in each radial bin. These results are shown in Fig.~\ref{fig:errors} (upper panel), together with the stacked surface brightness profile. Both profiles agree well within $R_{200c}$; however, at $R \gtrsim  R_{200c}$, the median profile drops significantly below the stacked profile, as expected for the case when the signal is dominated by individual bright regions spread across mostly negligible background. On the one hand, this highlights an advantage of using the stacked profile in this regime, while on the other, it shows that the stacked profile is not a representation of the typical cluster profile at these radii, but rather is a population-related probe of its own.}

\nl{In this regard, it is interesting to compare azimuthal variations within one cluster to variations within a population of clusters in the signal-dominated regime, where the stacked and median profile closely follow each other. The lower panel in Fig.~\ref{fig:errors} shows a comparison between uncertainties in the stacked surface brightness profile, which reflect variations across the four $90^\circ$ wedges, with the uncertainty on the median profile estimated as the standard deviation of the 136 individual wedge measurements (normalized according to the sample size). The similar amplitudes of the derived uncertainties indicate that cluster-to-cluster variations are comparable in amplitude to azimuthal (wedge-to-wedge) variations. 
We defer more detailed analysis involving rotation to match certain selected directions, e.g. connected to the orientation of the cluster's brightest galaxy, intracluster light, or surrounding large-scale structure, to the future work.}

\section{Conclusions}\label{sec:conclusions}

Recent all-sky observations with the \textit{SRG}/eROSITA telescope allowed us to construct the average soft X-ray surface brightness profile of massive clusters up to unprecedented radial distances and extremely low levels of the surface brightness. We take advantage of the identically computed predictions from the \texttt{Magneticum} suite of cosmological simulations and consider the stacked soft X-ray ($0.3 - 2.3$ keV) image for a sample of 34 most massive galaxy clusters. This procedure involves proper image rescaling and convolution of the predicted photon spectra with the eROSITA-like response function, as well as image filtering by excluding clusters' substructures and surrounding haloes.
The resulting profile turns out to be in good agreement with the measured averaged profile of clusters from the all-sky survey by the \textit{SRG}/eROSITA up to the radius $R_{200m} \approx 3\times R_{500}$; \hlcolor{however, a clear difference between the observed and the simulated profiles is present in the very central region ($\lesssim 0.4 \times R_{500}$), probably due to the implementation of the AGN feedback in simulations. 
Beyond $R_{200m}$, the stacked surface brightness profile reaches a constant radially uniform level (local background), whose origin is likely unrelated to the emission from the cluster. Moreover, the surface brightness profile derived directly from the gas density of the massive clusters from \texttt{Magneticum} reaches almost the same level, meaning correct accounting for the contribution of structures in the outskirts.
However, at these radii the local background emission becomes dominated by noise, and the mean signal from the cluster drops by orders of magnitude relative to its level. This suggests that proper detection of the component corresponding to cluster emission will remain a very challenging task even with larger samples in the future.} 


\section*{Data availability}

The catalogue of \texttt{Magneticum} clusters and the corresponding photon lists are publicly available online\footnote{\href{http://www.magneticum.org/data.html\#XRAY}{\texttt{http://www.magneticum.org/data.html\#XRAY}}}. The extracted ({un}filtered and filtered for bright substructures) cluster images and Radially-Resolved Clusters Spectral Database are available online\footnote{ \href{https://github.com/pi4imu/RRCS_DB}{\texttt{https://github.com/pi4imu/RRCS\_DB}}}.

\section*{Acknowledgements }

\rev{The authors are thankful to the anonymous referee for helpful comments that improved the quality of this work} and to M. Angelinelli for providing the gas density profiles of galaxy clusters from the \texttt{Magneticum} simulations.
AK and NL acknowledge financial support from the Russian Science Foundation (grant no. 25-22-00470). IK and KD acknowledge support by the COMPLEX project from the European Research Council (ERC) under the European Union’s Horizon 2020 research and innovation program grant agreement ERC-2019-AdG 882679.

\appendix

\section{Impact of the PSF smearing and redshift distribution of the clusters} \label{app:psf}

\begin{figure}[h]
  \centering
  \includegraphics[width=\linewidth,trim=0cm 0.5cm 0cm 0cm]{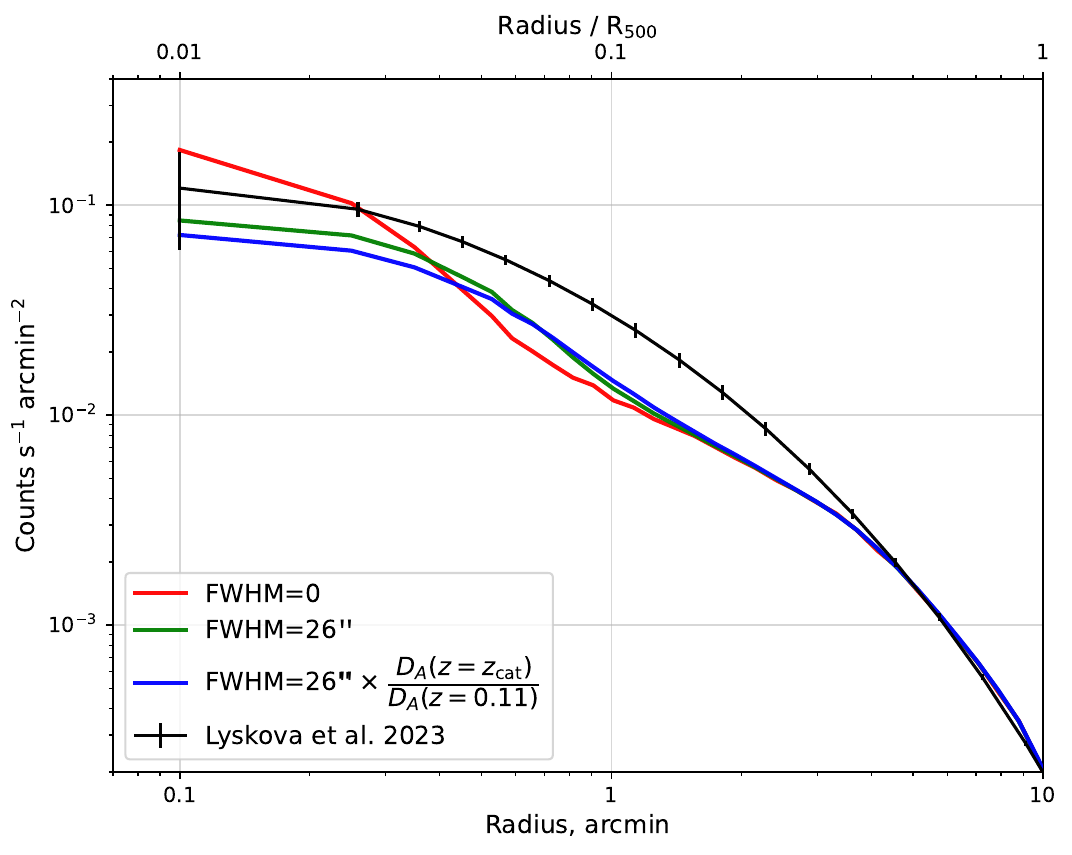}
  \caption{\hlcolor{Surface brightness profiles for the stacked images with different PSF: no smearing (red); smoothed with the FoV-averaged PSF with half-energy width corresponding to FWHM=$26''$ (green); and smoothed with a FWHM=$26''$, but rescaled as if each cluster were at $z=0.11$, the median redshift for the sample from \Ly~ (blue).}
 }
  \label{fig:psf}
\end{figure}

To illustrate the effect introduced by the\textit{SRG}/eROSITA PSF, we compare the surface brightness profiles for three different stacked images. For each individual image, we consider three different smoothing kernels. First, as if there is no PSF effect: FWHM=$0''$. Second, with FWHM=$26''$ (this setup is considered throughout this paper). Third, with FWHM=$26''$, but additionally multiplied by the factor $\frac{D_A(z=z_\mathrm{cat})}{D_A(z=0.11)}$ (where $D_{A}$ is angular diameter distance, $z_{cat}$ is cluster redshift from the catalogue and $z=0.11$ is median redshift of the sample from \Ly). The comparison is shown in Figure~\ref{fig:psf}.

\section{Manual filtering} \label{app:manual}

\begin{figure}
  \centering
  \hspace{0.1cm}
  \includegraphics[width=0.95\linewidth, trim = 0cm 0cm 17cm 0cm, clip]{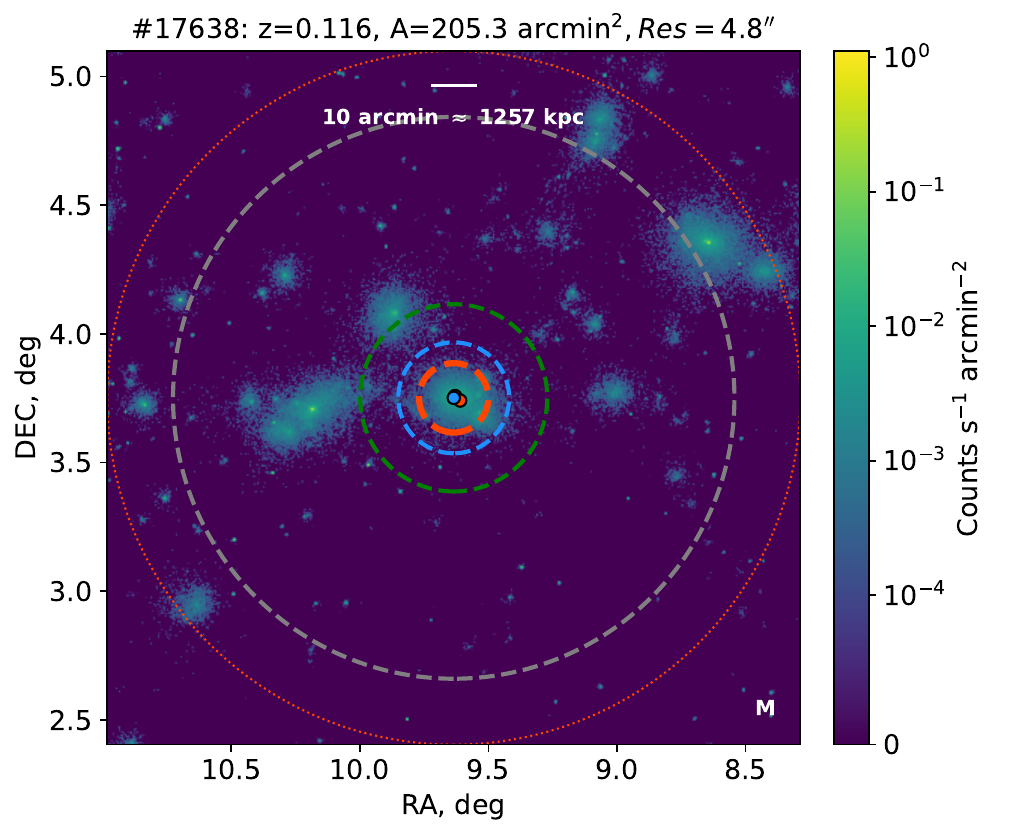}
  \hspace{2cm}
  \includegraphics[width=0.95\linewidth, trim = 17cm 0cm 0cm 0cm, clip]{fig6-17638-yes.pdf}
  \hspace{2cm}
  \includegraphics[width=0.95\linewidth, trim = 17cm 0cm 0cm 0cm, clip]{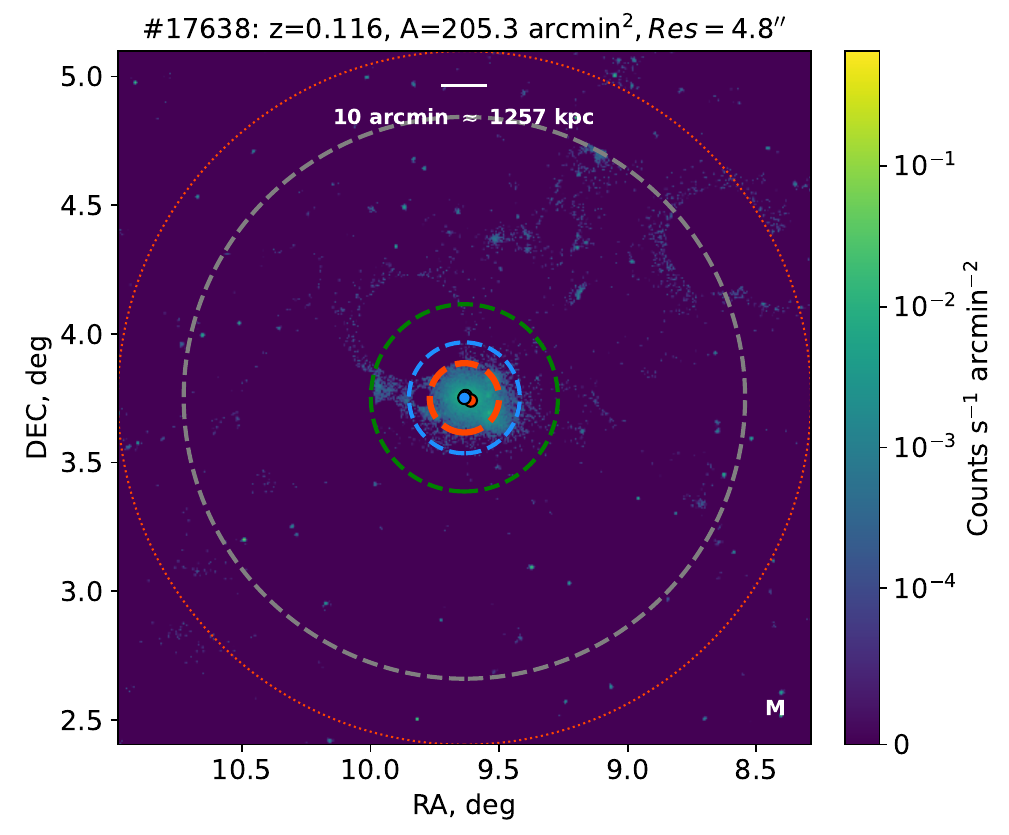}
  \caption{An example of the simulated cluster X-ray image and its additional filtering. \textit{Top:} the original image. \textit{Center:} the same image after automatic masking of catalogued haloes. \textit{Bottom:} the same image after additional manual filtering from structures that are missing in the halo catalogue (see text).}
  \label{fig:manual}
\end{figure}

Several massive clusters from the lightcone catalogue were additionally filtered manually. Clusters $\#1819, \#7308, \#11141, \#14857$ have bright substructures inside $R_{200m}$; $\#4613, \#17421$ have bright satellites on periphery ($> R_{ta}$) and $\#17638$ has unfiltered margins of masked substructures. For example, the image of cluster \#17638 is shown in Fig.~\ref{fig:manual}: residual emission east of the cluster centre, \hlcolor{i.e. relatively bright outskirts of surrounding haloes that were filtered and excluded automatically,} was masked manually.

\section{Average surface brightness profile from gas density profiles and clumpiness estimation} \label{app:a22}

\begin{figure}[h]
  \centering
  \includegraphics[width=\linewidth]{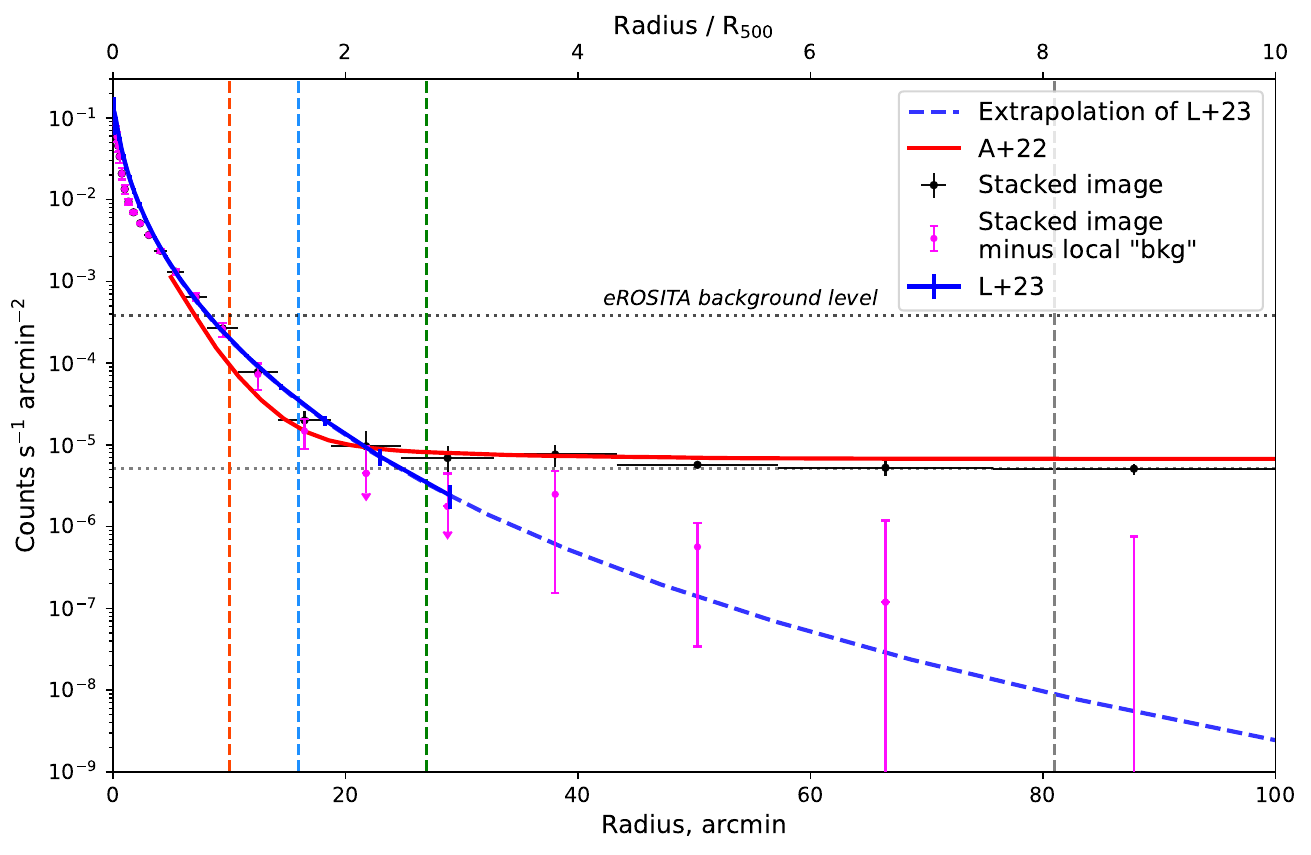}
  \hspace*{0.07cm}
  \includegraphics[width=0.99\linewidth,trim=0cm 0.5cm 0cm 0.8cm]{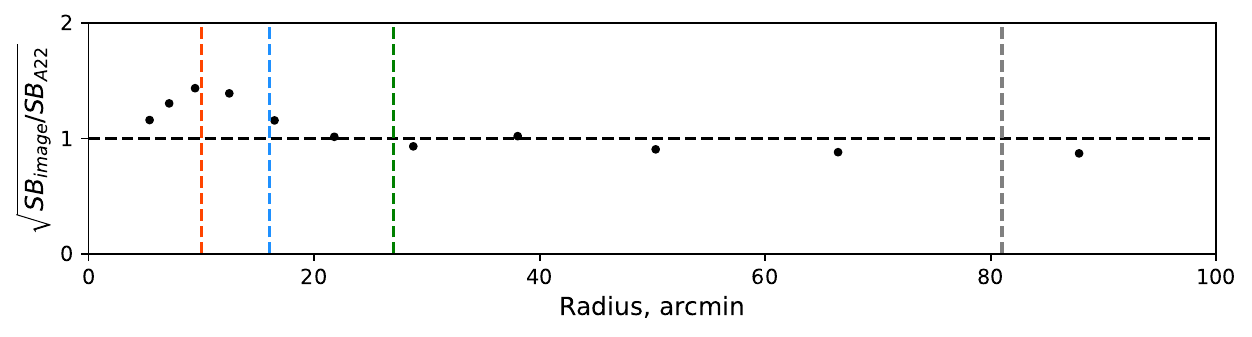}
  \caption{Comparison of the directly calculated surface brightness profile with the one derived from the averaged density profile.  \textit{Top:} same as for Fig.~\ref{fig:profiles_linear}, but with the solid red line showing the surface brightness derived directly from gas density of \texttt{Magneticum} galaxy clusters \citep{2022A&A...663L...6A}. \textit{Bottom:} square root of ratio of the stacked surface brightness profile to the profile derived from gas density, which might be used as a proxy of the gas clumpiness.
  }
  \label{fig:profiles_a22} 
\end{figure}

{To ensure that our \hlcolor{mock observations} properly capture the \hlcolor{overall} contribution of the outskirts \hlcolor{to the total} emission, we compare our stacked surface brightness profile with the one derived directly from the gas density of \texttt{Magneticum} clusters considered in the earlier studies \citep{2022A&A...663L...6A}. The procedure is carried out as follows:} we compute the mean gas density profile $n_{gas}^{mean}(r)$ of 5 most massive clusters from \cite{2022A&A...663L...6A} (see their Fig.~1) and then integrate its square along the line-of-sight using Abel transform in order to obtain the two-dimensional projected surface brightness profile:

\begin{equation}
S_X(R) = 2 \int_{R}^{R_{max}} \frac{r}{\sqrt{r^2-R^2}} \cdot \Bigg[ \frac{n_{gas}^{mean}(r)}{N} \Bigg]^2  dr,
\end{equation}

where $R_{max}=10 \times R_{500} = 100 \ {\rm arcmin}$ and $N = 0.014 \ {\rm cm}^{-3}$ is the normalization factor derived in \Ly~for the relevant \textit{SRG}/eROSITA observational parameters (see their Eq.~3 and Appendix~C there), which have identical values in this work. \hlcolor{The obtained surface brightness profile is shown in} Fig.~\ref{fig:profiles_a22} \hlcolor{in solid red line, demonstrating good agreement with our result over entire available radial range. This result implies that} \hlcolor{the analysis from} \cite{2022A&A...663L...6A}, \hlcolor{despite not introducing observational-like artifacts, still includes contributions from the local structures (outskirts of unresolved sub-structures and neighbouring structures, and unresolved gas structures).}

\hlcolor{ICM can be very inhomogenous (i.e. "clumpy"), which leads to overestimation of the gas density from the X-ray analysis} \citep{2011ApJ...731L..10N, 2013MNRAS.428.3274Z, 2013MNRAS.432.3030R, 2021A&A...653A.171A}. To estimate the level of clumpiness (i.e., factor $C$, where $C=\sqrt{\langle n_{gas}^2 \rangle / \langle n_{gas} \rangle^2}$ with $\langle .. \rangle$ meaning volume averaging) for the stacked cluster\hlcolor{s}, we compute the square root of the ratio of the stacked surface brightness profile to the profile derived directly from the gas density; the resulting ratio is $\leq 1.7$ (\hlcolor{see bottom panel of} Fig.~\ref{fig:profiles_a22}), indicating that a relatively mild level of gas clumping is required for consistency.

\section{Comparison of the profiles for different cluster samples}
\label{app:bigsample}

\begin{figure}[ht]
  \centering
  \includegraphics[width=\linewidth,trim=0cm 0.5cm 0cm 0cm]{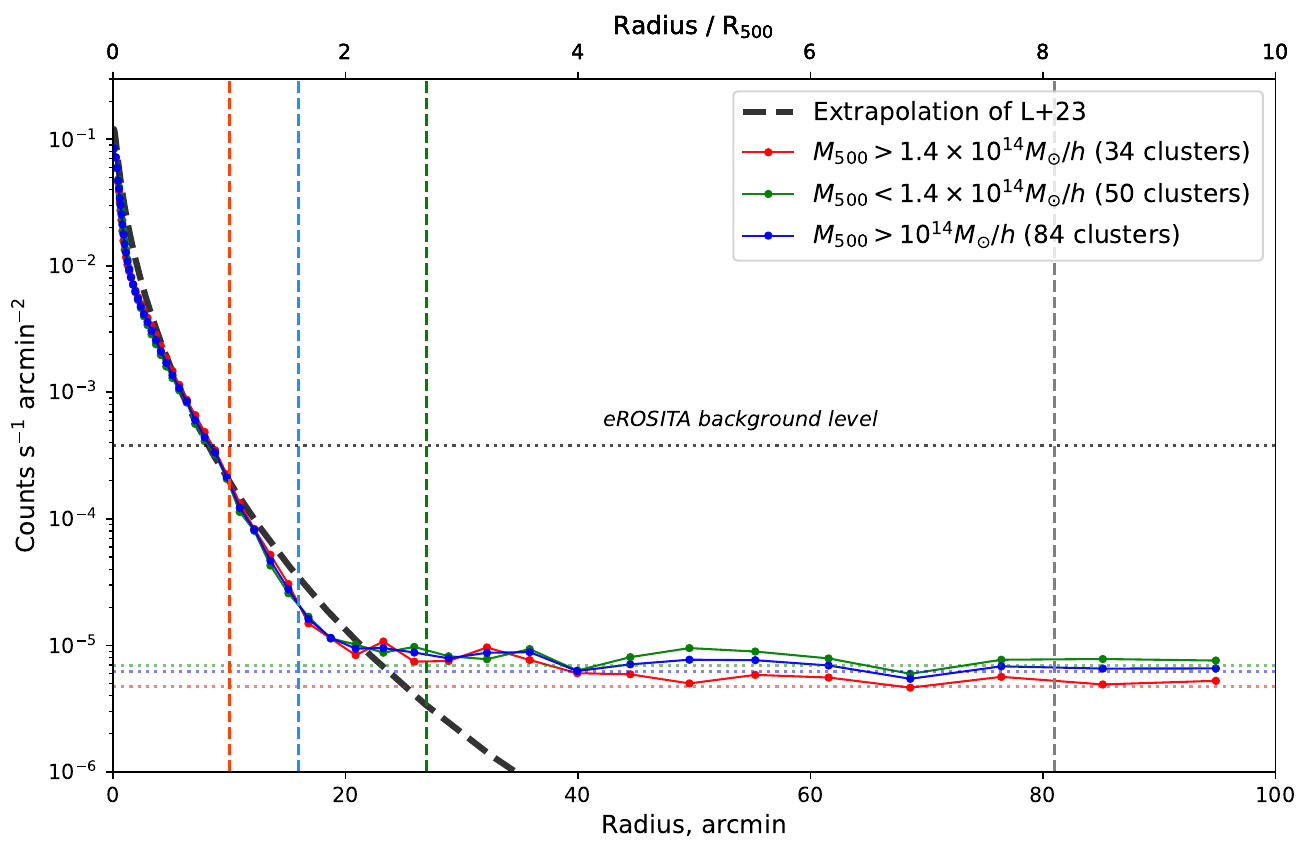}
  \caption{\hlcolor{Radial surface brightness profiles of the stacked image in the 0.3-2.3 keV band for different \texttt{Magneticum} cluster samples: the full sample (blue); massive clusters analyzed in this paper (red); the remaining subsample (green).}}
  \label{fig:profiles_linear_all}
\end{figure}

Fig.~\ref{fig:profiles_linear_all} shows a comparison of the stacked profile for 34 clusters with $M_{500} > 1.4 \times 10^{14} M_{\odot}/h$  analysed in this work with that for all 84 massive clusters in the lightcone (with $M_{500} > 10^{14} M_{\odot}/h$) and that for the 50 remaining clusters. \hlcolor{The profiles agree up to the value $2 \times R_{500} = 20$ arcmin; moreover, beyond this radius, they show nearly identical behavior, suggesting an approximately self-similar surface brightness and, consequently, gas distribution over the mass range of the subsamples considered.}

\section{\revv{Stacked vs. median profiles and comparison of uncertainties}}
\label{app:errors}

\begin{figure}[ht]
  \centering
  \includegraphics[width=0.99\linewidth,trim=0cm 0.88cm 0cm 0cm]{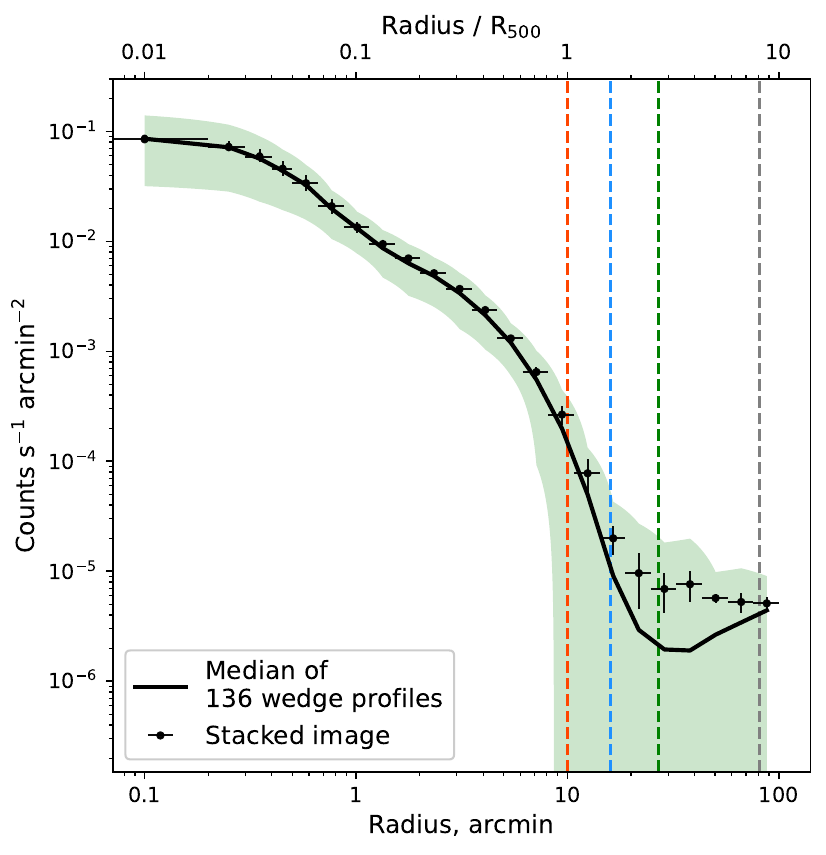}
  \hspace*{-0.22cm}
  \includegraphics[width=1.01\linewidth,trim=0cm 0.3cm 0cm 0.23cm, clip]{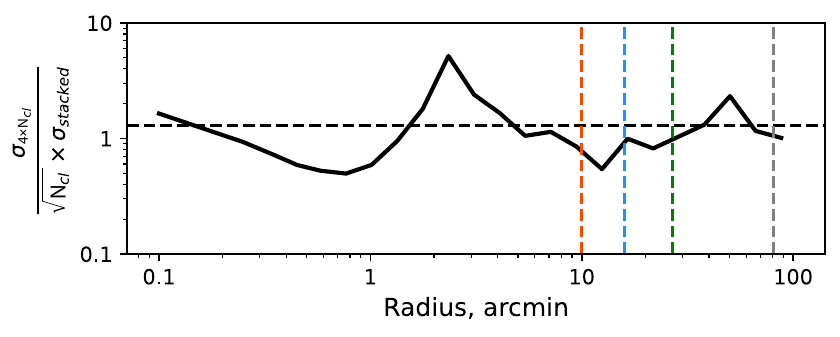}
  \caption{
  \revv{Comparison of the stacked and median surface brightness profiles and uncertainties in the measured profiles. \textit{Top:} the stacked image profile (black points with error bars) and median profile from four 90-degree wedges for each cluster image ($4\times34=136$ wedges in total) with the $1\sigma$ uncertainty indicated by the green-shaded region.
  \textit{Bottom:} $1\sigma$ uncertainty values for 136 wedges (normalized by the square root of the sample size) divided by surface brightness errors from the stacked image, as a function of radius. The horizontal dashed line shows the mean value of this ratio, which is close to unity. }
   }
  \label{fig:errors}
\end{figure}

\revv{Fig.~\ref{fig:errors} illustrates of the results discussed in Section~\ref{sec:stvsmed}. In the upper panel, we compare the surface brightness profile obtained from the stacked image with the median of the 136 wedge profiles: there is a good agreement between both profiles within $R_{200c}$, but at $R_{200m}$ the median profile has a significant drop relative to the stacked profile. In the lower panel, we compare the standard deviation of surface brightness values across the four wedges of the stacked image with the standard deviation of the 136 individual wedge measurements (divided by the square root of 34), and these two estimates are similar.}

\bibliographystyle{elsarticle-harv} 
\bibliography{bibfile}

\end{document}